\documentclass[Physsubmission, Phys]{SciPost}

\hypersetup{
    colorlinks,
    linkcolor={red!50!black},
    citecolor={blue!50!black},
    urlcolor={blue!80!black}
}

\usepackage[bitstream-charter]{mathdesign}
\DeclareSymbolFont{usualmathcal}{OMS}{cmsy}{m}{n}
\DeclareSymbolFontAlphabet{\mathcal}{usualmathcal}

\begin{document}

\begin{center}{\Large \textbf{
Multiplicity dependence of charged-particle jet production in pp collisions at 13 TeV with ALICE\\
}}\end{center}

\begin{center}
Debjani Banerjee\textsuperscript{1$\star$} for the ALICE Collaboration
\end{center}

\begin{center}
{\bf 1} Bose Institute, Kolkata 700091, India
\\
* banerjee.debjani@cern.ch
\end{center}

\begin{center}
\today
\end{center}


\definecolor{palegray}{gray}{0.95}
\begin{center}
\colorbox{palegray}{
  \begin{tabular}{rr}
  \begin{minipage}{0.1\textwidth}
    \includegraphics[width=23mm]{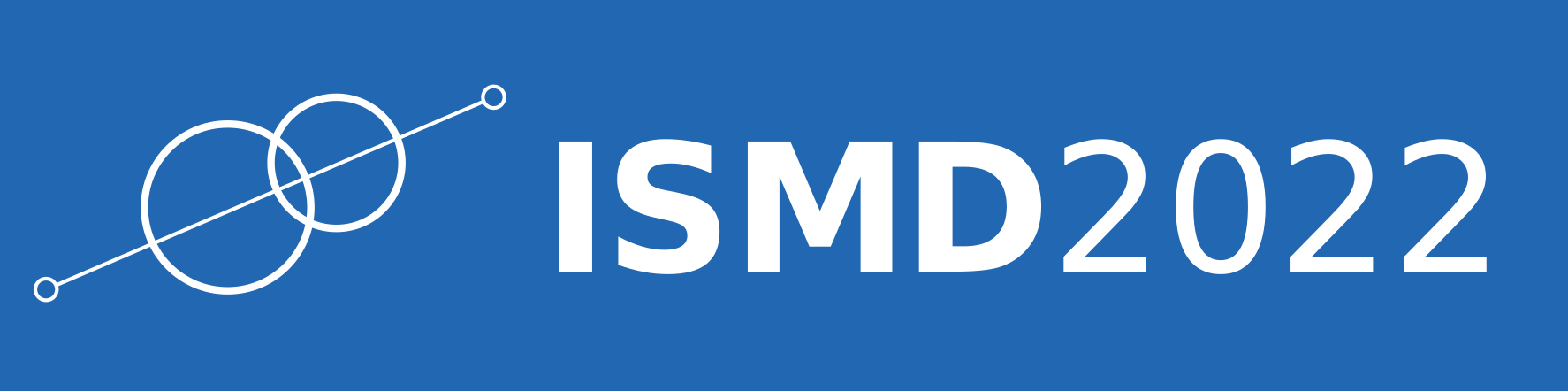}
  \end{minipage}
  &
  \begin{minipage}{0.8\textwidth}
    \begin{center}
    {\it 51st International Symposium on Multiparticle Dynamics (ISMD2022)}\\ 
    {\it Pitlochry, Scottish Highlands, 1-5 August 2022} \\
    \doi{10.21468/SciPostPhysProc.?}\\
    \end{center}
  \end{minipage}
\end{tabular}
}
\end{center}

\section*{Abstract}
{\bf

Measurements of jet production and jet properties in pp collisions
provide a test of perturbative quantum chromodynamics (pQCD) and form
a baseline for similar measurements in heavy ion (A--A) collisions.
In this contribution, we report recent ALICE measurements of
charged-particle jet production and intra-jet properties,
including mean charged-constituent multiplicity and fragmentation
distribution for leading jets, in minimum bias pp collisions at
$\sqrt{s}$ = 13\,TeV. In addition, the event multiplicity dependence
of jet production and jet properties in pp collisions at $\sqrt{s}$ =
13\,TeV will be presented.
Results will be compared with theoretical model predictions.

\vspace{10pt}
\noindent\rule{\textwidth}{1pt}
\tableofcontents\thispagestyle{fancy}
\noindent\rule{\textwidth}{1pt}
\vspace{10pt}

\section{Introduction}
\label{sec:intro}
The hard-scattered (high $p_{\rm T}$) partons produced in high-energy collisions fragment into a collimated shower of final-state particles, known as jets. Jets are a multipurpose tool to explore various properties of the strong interaction and provide an important benchmark for perturbative quantum chromodynamics (pQCD) predictions. Measurement of jet production and their properties are sensitive to the details of parton shower and hadronization. In addition, studying jet properties 
in small collision systems like proton-proton (pp) collisions, especially in high multiplicity events, is important in order to look for the onset of quark-gluon plasma (QGP)-like effects as a function of particle multiplicity. In this work, we present the multiplicity dependence of charged-particle jet production, measurements of intra-jet properties for leading jets, the mean charged-particle multiplicity, $\left<N_{\rm ch}\right>$,  and fragmentation functions, $z^{\rm ch} = {p_{\rm T}^{\rm particle}}/{p_{\rm T}^{\rm jet,ch}}$ in pp collisions at $\sqrt{s}$ = 13\,TeV with ALICE. Results are compared to Monte Carlo generators which help to constrain the models and provide inputs to tune.

\section{Analysis details}
For the measurement of jet properties, charged particles with $p_{\rm T} >$ 0.15 GeV/$c$ with pseudorapidity $|\eta| <$ 0.9 and azimuthal angle 0 $ < \varphi <$ 2$\pi$ are considered. Jets are constructed from these selected charged particles with FastJet 3.2.1~\cite{FastJet} using the anti-$k_{\rm T}$ algorithm for jet resolution parameter $R$ = 0.4. High multiplicity (HM) events are selected using the HM trigger condition which requires the sum of V0A and V0C~\cite{V0} amplitudes to be more than 5 times the mean MB signal. The underlying event contribution is estimated using the perpendicular-cone method~\cite{jetpp7TeV} and subtracted on a statistical basis after unfolding. A 2-D Bayesian unfolding technique~\cite{Bayesian} implemented in RooUnfold~\cite{Roounfold} package is applied to correct for the instrumental effects. 
The dominant systematic uncertainty in this analysis is the uncertainty on the tracking efficiency. The analysis procedure for the measurement of the multiplicity dependence of jet production in pp collisions at $\sqrt{s}$ = 13\,TeV is described in detail in Ref.~\cite{jetprodpp}. 

\section{Results and discussion}
\begin{figure}[h!]
\centering
\includegraphics[width=0.9\textwidth]{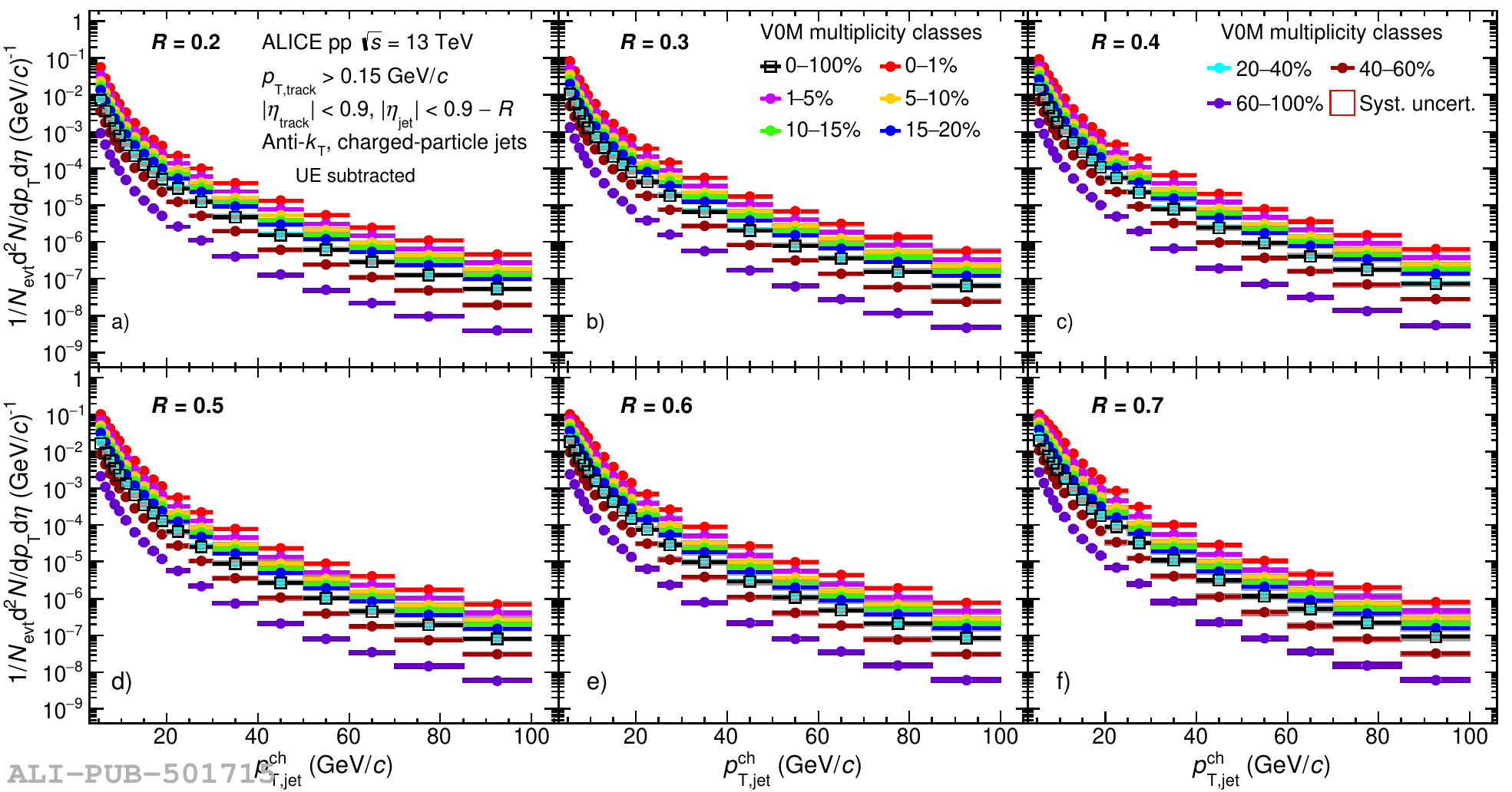}
\caption{Charged-particle jet yields in different multiplicity percentile intervals for resolution parameter $R$ varied from 0.2 to 0.7.}
\label{jetprod}
\end{figure}
\begin{figure}[h!]
	\centering
	\includegraphics[width=1.0\textwidth]{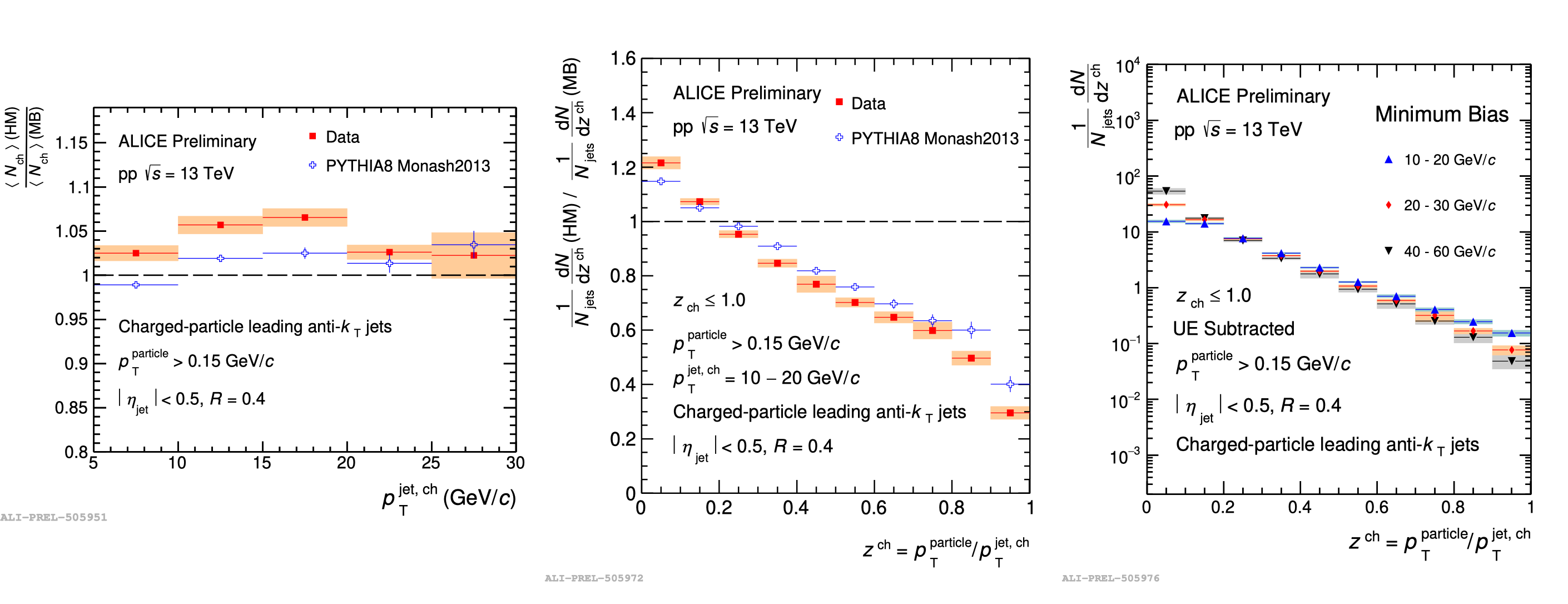}
	\caption{The ratio $\left<N_{\rm ch}\right>$ (HM)/$\left<N_{\rm ch}\right>$ (MB) as a function of jet $p_{\rm T}$ (left), the ratio of the $z^{\rm ch}$ distributions between HM and MB events (middle), and  $z^{\rm ch}$ distributions with different jet $p_{\rm T}$ ranges for MB events (right).}
	\label{jetprop}
\end{figure}
Figure~\ref{jetprod} shows the charged-particle jet production yields in various multiplicity percentiles as a function of jet $p_{\rm T}$ for jet resolution parameter $R$ varied from 0.2 to 0.7 in pp collisions at $\sqrt{s}$ = 13\,TeV. It is observed that the jet yield is higher (lower) in higher (lower) multiplicity classes. Furthermore, it is observed that the changes of these distributions in different multiplicity classes with respect to the minimum bias case are independent of the jet resolution parameter $R$ and have a weak $p_{\rm T}$ dependence, except for the very low $p_{\rm T}$ domain. This indicates that jet production changes with event activity, but the slope of the spectrum stays similar to the one measured in MB events~\cite{jetprodpp}.

The ratio $\left<N_{\rm ch}\right>$ (HM)/$\left<N_{\rm ch}\right>$ (MB) is shown as a function of leading $p_{\rm T}^{\rm jet,ch}$ in Fig.~\ref{jetprop} (left). The $\left<N_{\rm ch}\right>$ is larger for HM events and is qualitatively reproduced by PYTHIA8 Monash2013~\cite{pythia8} for $p_{\rm T}^{\rm jet,ch} <$ 20 GeV/$c$. Fig.~\ref{jetprop} (middle) shows the ratio of $z^{\rm ch}$ distributions obtained in HM and MB events. It is interesting to notice that the jet fragmentation is softer in HM events. Fig.~\ref{jetprop} (left) shows the $z^{\rm ch}$ distributions in three jet $p_{\rm T}$ domains for MB events. The charged-particle jet fragmentation scales with jet $p_{\rm T}$ except at highest and lowest $z^{\rm ch}$.
\section{Conclusion}
We have presented the multiplicity dependence of jet production and intra-jet properties in pp collisions at 13\,TeV with ALICE. Results are compared with the MC generators. We have observed that jet production changes with multiplicity, but the slope of the distributions remains similar to the one measured in MB events. For the intra-jet properties, a significant modification in jet fragmentation distributions is observed in HM events compared to MB events. We have also observed that the jet fragmentation is softer in the HM events. In the future, the modification of the fragmentation function will be studied at higher jet $p_{\rm T}$.

\section*{Acknowledgements}
Debjani Banerjee would like to acknowledge the DST INSPIRE research grant [DST/INSPIRE Fellowship/2018/IF180285] and the ALICE project grant [SR/MF/PS-02/2021-BI\,(E-37125)]. Significant part of computation for this work was carried out using the computing server facility at CAPSS, Bose Institute, Kolkata.







\end{document}